\documentclass[aps,twocolumn,showpacs,superscriptaddress,floatfix]{revtex4}
\usepackage[T1]{fontenc}
\usepackage{graphicx}
\usepackage{ae}
\begin{document}

\title{Experimental evidence for the breakdown of a Hartree-Fock
approach in a weakly interacting Bose gas}

\author{J.-B.~Trebbia}
\affiliation{Laboratoire Charles Fabry, CNRS et Université Paris 11,
91403 Orsay CEDEX, France}
\author{J.~Esteve}
\affiliation{Laboratoire Charles Fabry, CNRS et Université Paris
11, 91403 Orsay CEDEX, France} \affiliation{Laboratoire de
Photonique et de Nanostructures, CNRS, 91460 Marcoussis, France}
\author{C. I.~Westbrook}
\affiliation{Laboratoire Charles Fabry, CNRS et Université Paris
11, 91403 Orsay CEDEX, France}
\author{I.~Bouchoule}
\affiliation{Laboratoire Charles Fabry, CNRS et Université Paris
11, 91403 Orsay CEDEX, France}

\begin{abstract}
 We investigate  the physics underlying the presence of
a quasi-condensate in a nearly one
dimensional, weakly interacting trapped atomic Bose gas.
We show
that a Hartree Fock (mean-field) approach fails to predict the
existence of the quasi-condensate in the center of the cloud: the
quasi-condensate is generated by
interaction-induced correlations between atoms
and not by a saturation of the excited states.
Numerical calculations
based on Bogoliubov theory give an estimate of the cross-over
density in agreement with experimental results.
\end{abstract}

\maketitle

Since the first observation of the Bose-Einstein condensation in
dilute atomic gases, self consistent mean-field approaches (see
e.g.~\cite{griffinHF,giorgini97}) have successfully described most
experimental results. A good example is the critical temperature at
which the condensate
appears~\cite{shiftTcFabrice,shiftTcStringari,Holzmann1999}.
However, the approximate treatment made in these theories about
particle correlations can not capture all the subtle aspects of the
many body problem and the success of these theories relies on the
weakness of interactions in dilute atomic gases ($\rho a^3 \ll 1$
where $\rho$ is the atomic density and $a$ the scattering length for
three-dimensional (3D) gases) and  also on the absence of large
fluctuations (Ginzburg criterion)~\cite{LandauGinzbourgCriterium}.
In a 3D Bose gas, this latter criterion is not fulfilled for
temperatures very close to the critical
temperature~\cite{shiftTcStringari} and deviations from the
mean-field theories are expected. For a given atomic density, the
critical temperature is calculated to be slightly shifted towards a
larger value as compared to the mean-field
prediction~\cite{shiftTcLaloe,shiftTcMoore,shiftTcSvistunov} but
this discrepancy is still beyond the precision of current cold atom
experiments~\cite{shiftTcpiege,shiftTcFabrice}.

 In mean-field theory, a Bose gas above condensation is described by a
Hartree-Fock (HF) single fluid in which atomic interactions are
taken into account only {\it via} a mean-field  potential
$V_{\rm{m.f.}}=2g\rho$\cite{lesHouches1999Castin,StenholmHF} where
$g$ is the coupling constant and $\rho$ the atomic density.
 Interaction-induced atomic correlations are neglected and the
gas is modeled as a group of non-interacting bosons that experience
the self-consistent potential $V_{\rm{m.f.}}$. As for an ideal Bose
gas, the two particle correlation function at zero distance
$g^{(2)}(0)$ is 2 (bunching effect) \cite{LandauGinzbourgCriterium}.
This HF single fluid description holds until the excited state
population saturates, which is the onset of Bose condensation.

In this paper, we present measurements of density profiles of a
degenerate Bose gas in a situation where the trap is very elongated
and the temperature of the cloud is close to the transverse ground
state energy. For sufficiently low temperatures and high densities,
we observe the presence of a quasi-condensate
\cite{Phasefluctu_Petrov} at the center of the cloud. Using the
above HF theory, we show that a gas with the same temperature and
chemical potential as the experimental data is \textit{not} bose
condensed. Thus, a mean-field approach doesn't account for our
results and the quasi-condensate regime is not reached \textit{via}
the usual saturation of the excited states.
 We emphasize that this failure of mean-field theory happens
in a situation where the gas is far from the strong interaction
regime, which in one-dimension (1D), corresponds to the
Tonks-Girardeau gas limit
\cite{Kheru2003,Bose1D_Lininger,Phasefluctu_Petrov}, and where mean
field theory also fails.
 To our knowledge, this is the first demonstration of the breakdown
of a Hartree-Fock approach in the weakly interacting
limit.

 We attribute this failure of the mean-field theory to
the nearly 1D character of the gas. It is well known that a 1D
homogeneous ideal Bose gas does not experience Bose Einstein
condensation in the thermodynamic limit.
 On the other hand, in the
 presence of repulsive interactions in the weakly interacting regime,
as the linear density increases one expects a smooth cross-over from
an ideal gas regime where $g^{(2)}(0)\simeq 2$ to a quasi-condensate
regime where $g^{(2)}(0)\simeq
1$\cite{Kheru2003,Quasibec_Castin,Castin1Dclass}. The HF approach
fails to describe this cross-over:~as for an ideal gas, the thermal
fluid does not saturate and $g^{(2)}(0)=2$ for any density. The
above results also hold for a 1D harmonically trapped gas at the
thermodynamic limit:~no saturation of excited states occurs for an
ideal gas \cite{Bagn91} and the gas smoothly enters the
quasi-condensate regime when the peak density
increases\cite{Phasefluctu_Petrov,Kheru2005}.

In the experiment presented here, the gas is neither purely 1D nor
at the thermodynamic limit: a few transverse modes of the trap are
populated and a condensation phenomenon due to finite size effects
might be expected \cite{Ketterlecond1D}. However, we will show that,
as in the 
the scenario  discussed above,  the
gas undergoes a smooth cross-over to the quasi-condensate regime
without saturation of the excited states.
An estimation of the cross-over density using
a three-dimensional Bogoliubov calculation is
in agreement with experimental data.

\begin{figure*}[t]
\centerline{
 \includegraphics[]{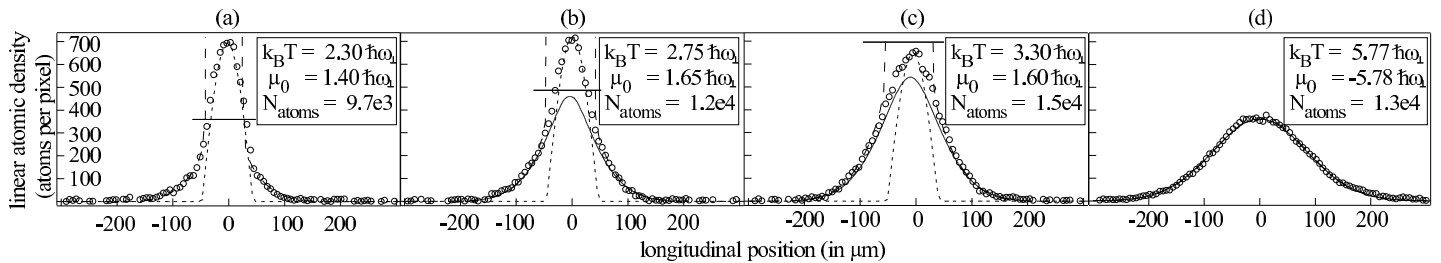}
} \caption{ \textit{In situ }longitudinal distributions for
different temperatures (circles).  Long
dashed lines:~ideal Bose gas profile at the shown temperature $T$ and chemical potential $\mu_0$
obtained from a fit to the wings. Solid lines:~profiles
obtained in the Hartree-Fock approximation for the same $T$ and $\mu_0$.
 Short dashed lines:~quasi-condensate
profiles with the same peak density as the experimental
data.
 Horizontal solid lines:~cross-over density
 estimated
using a Bogoliubov calculation (see text).} \label{figure_resultats}
\end{figure*}

 The experimental setup is the same as \cite{esteve2006}.
 Using a Z-shaped wire on an atom chip \cite{Reic99}, we produce an
 anisotropic trap, with a transverse frequency of $\omega_{\perp}/(2 \pi)=  2.75$~kHz
and a longitudinal frequency of $\omega_{z}/(2 \pi)= 15.7$~Hz.
 By evaporative cooling, we obtain a few thousand $^{87}$Rb atoms
 in the $|F=2,m_F=-2\rangle$ state at a temperature of a few times $\hbar \omega_{\perp}/k_B$.
Current-flow deformations inside the micro-wire, located 150 $\mu$m
below the atoms, produce a roughness on the longitudinal potential
\cite{Esteve2004}. The observed atomic profiles are smooth (see
Fig.~\ref{figure_resultats}), which shows that this roughness is
small and we neglect it in the following.

 The longitudinal profile of the trapped gases are recorded using
{\it in situ} absorption imaging  as in \cite{esteve2006}.
 The probe beam intensity is about 20\% of
the saturation intensity and the number of atoms contained in a
pixel of a CCD camera is deduced from the formula $N_{\rm
at}=(\Delta^2/\sigma) \rm{ln}({I_2}/{I_1})$, where
$\Delta=6.0~\mu$m is the pixel size, $\sigma$ the effective cross
section, and $I_1$ and $I_2$ the probe beam intensity respectively
with and without atoms. The longitudinal profiles are obtained by
summing the contribution of the pixels in the transverse
direction. However,  when the optical density is large and the
density varies on a scale smaller than the pixel size the above
formula underestimates the real atomic density \cite{esteve2006}.
In our case, the peak optical density at resonance is about $1.5$,
and this effect cannot be ignored. To circumvent this problem, we
decrease the absorption cross section by detuning the probe laser
beam from the $F=2\rightarrow F'=3$ transition by $\delta=9$~MHz.
 We have checked that for larger detunings, the normalised profile remains
identical to within 5\%.
 For the detuning $\delta$, the lens effect due to the real part
of the atomic refractive index is calculated to be small enough so
that all the refracted light is collected by our optical system
and the profile is preserved.

To get an absolute measurement of the linear density, we need the
effective absorption cross section $\sigma$. We find $\sigma$ by
comparing the total absorption of {\it in situ} images taken with a
detuning $\delta$ with the total number of atoms measured on images
taken at resonance after a time of flight long enough (5~ms) that
the optical density is much smaller than 1. In these latter images,
the probe beam is $\sigma^+$ polarized and the magnetic field is
pointing along the probe beam propagation direction so that
the absorption cross section is
$3\lambda^2/2\pi$. We obtain for the {\it in situ} images taken at a
detuning $\delta$, an effective absorption cross section $(0.24\pm
0.04)\times 3\lambda^2/2\pi$.
 For samples as cold as that in Fig. \ref{figure_resultats} (a),
the longitudinal profile is expected to be unaffected by the time of
flight and we checked that the profile is in agreement with that
obtained from {\it in situ} detuned images within 5\%.

 Averaging over 30 measured profiles, we obtain a
relative accuracy of about 5\% for the linear density. A systematic
error of about 20\% is possible due to the uncertainty in the
absorption cross section.

 In Fig.~\ref{figure_resultats}, we plot the longitudinal
density profiles of clouds at thermal equilibrium for different
final evaporating knives obtained from {\it in situ} images.
 We have compared the density profiles with the expected quasi-condensate density
profile with the same peak density. This is obtained using the
equation of state of the longitudinally homogeneous gas $\mu=\hbar
\omega_{\perp}\sqrt{1+4na}$ \cite{Gerbier:771}, and the local
chemical potential $\mu (z)= \mu_0-1/2m\omega_z^2z^2$, where $n$ is
the linear atomic density, and $\mu_0$ the chemical potential at the
center of the cloud. For the two colder clouds (graphs $(a)$ and
$(b)$) in Fig.~\ref{figure_resultats}, we observe a good agreement
between the central part of the experimental curves (circles) and
the quasi-condensate profile (short dashed lines) which indicates
that the gas has entered the quasi-condensate regime. In
\cite{esteve2006}, we observed the inhibition of density
fluctuations expected in this regime.

\begin{figure}
{\includegraphics{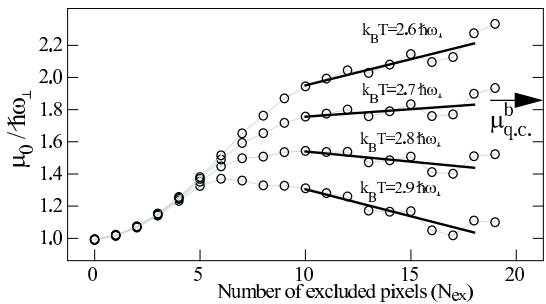}} \caption{ Chemical potential $\mu_0$,
obtained by fitting the wings of profile (b) to an ideal Bose gas
distribution, as a function of the number of excluded pixels
$N_{\rm ex}$ on either side of the distribution, for  four
different trial temperatures.
 The arrow indicates
$\mu_{\rm q.c.}=\hbar\omega_{\perp}\sqrt{1+4n_0a}$, which is the
chemical potential found from the measured peak linear density
$n_0$. } \label{figure_fit}
\end{figure}

 For a given atomic density profile, we extract the temperature and chemical potential of the data
from  a fit to an ideal Bose gas distribution. For cold clouds, the
ideal Bose gas model should fail in the central part of the cloud
where interactions are not negligible. We therefore fit only the
wings of the profile by excluding a number of pixels $N_{\rm ex}$ on
either side of the center of the profile.  We fit the longitudinal
atomic distribution with only the chemical potential $\mu_0$ as a
free parameter for different trial temperatures $T$. As seen in
Fig.~\ref{figure_fit}, $\mu_0(T,N_{\rm ex})$ is approximately linear
in $N_{\rm ex}$ for $N_{\rm ex}>10$.
 The cloud's temperature is the one for which  $\mu_0(T,N_{\rm ex})$
is independent of $N_{\rm ex}$. The fluctuations around the straight
line for $N_{\rm ex}>10$, seen in Fig.~\ref{figure_fit}, is mainly
due to the potential roughness. These fluctuations contribute about
for 5\% to the uncertainty of $T$ and $\mu_0$. The determination of
$\mu_0$ however, is primarily limited by the uncertainty in the
total atom number. For profile (b), we find $T_b=(2.75 \pm 0.05
)\hbar \omega_{\perp}/k_B$ and $\mu_0^b=(1.65\pm 0.5) \hbar
\omega_{\perp}$. For the profile (a), the chemical potential and the
temperature are not very accurate because the wings are too small.
This profile is not analyzed in the following.

 Another method to
deduce the chemical potential is to use the peak atomic density and
the formula $\mu=\hbar \omega_{\perp}\sqrt{1+4na}$,  assuming the
gas is well inside the quasi-condensate regime at the center of the
cloud. For graph $(b)$, we obtain  $\mu_{\rm q.c.}^{b} = (1.8 \pm
0.2)\ \hbar\omega_{\perp}$, which is consistent with the value
obtained from fits of the wings of the distribution.

We now compare the experimental density profiles with the
longitudinal atomic density $n(z)$ given by the Hartree Fock theory
for the same temperature and chemical potential. To do the
calculation, we also assume the relative population of the ground
state is negligible (no Bose-Einstein condensation) and the
quantization of the longitudinal eigenstates is irrelevant. In this
so called local density approximation (LDA), the linear atomic
density $n(z)$ is identical to the linear atomic density
$n_h(\mu(z),T)$ of a thermal Bose gas trapped in the radial
direction and longitudinally untrapped at a chemical potential
$\mu(z)=\mu_0-m\omega_z^2z^2/2$.

 To obtain $n_h(\mu,T)$, we need to compute the self-consistent
three-dimensional atomic density $\rho$ which is the thermodynamic
distribution of independent bosons that experience the self
consistent HF Hamiltonian\cite{lesHouches1999Castin,StenholmHF}
\begin{equation}
H_{HF}=\frac{p_z^2}{2m}+H_{\rm kin}+H_{\rm harm}+2g\rho(r),
\label{eq.HF}
\end{equation}
where   $H_{\rm kin}$ is the transverse kinetic energy term, $H_{
\rm harm}$ the transverse harmonic potential,  $r$ is the radial
coordinate and $g=4\pi\hbar^2a/m$ where $a$ is the Rb$^{87}$
scattering length. The simplest approach to obtain $\rho(r)$ is to
use an iterative method, starting from $\rho(r)=0$. For each
iteration, we numerically diagonalize the transverse part of
$H_{HF}$ to deduce the new thermal atomic density distribution. When
the interactions become too strong (linear density larger than 320
atoms per pixels for the temperature $T_b$)
 this algorithm does not converge.
In this case, we use a more time consuming method based on a
minimization algorithm. We use the trial function $\rho_{\rm trial}
(r) =\sum c_{2p} H_{2p}(r)e^{-r^2/2r_0^2} $where $H_{2p}(r)$ are
Hermite polynomials, and $0\leq\! p\!\leq 3$.
 We find $r_0$ and the four $c_{2p}$ coefficients by minimizing
$ \xi=\int_{0}^\infty r(\rho_{\rm
trial}^{'}(r)-\rho_{\rm trial}(r))^2dr/\int_{0}^\infty r\rho_{\rm
trial}^2(r)dr$, where $\rho_{\rm trial}^{'}(r)$ is the thermodynamic
equilibrium atomic density for $H_{HF}[\rho_{\rm trial} (r)]$.
 We find $\xi$ less
than $10^{-4}$ meaning that our 5 parameter model describes the
transverse Hartree Fock profile well.
 The linear density $n_h=\int 2\pi r \rho_{\rm trial}(r)dr$ is identical to
$n'_h=\int 2\pi r \rho_{\rm trial}'(r)dr$ within $0.5\%$. In the
domain where both methods are valid, we also check that they give
the same result.

 Figure~\ref{figure_resultats} compares the
longitudinal profiles obtained with the HF calculation
with the experimental data and the ideal gas profile for the graphs
$(b),(c)$ and $(d)$.
 The HF profile for the hottest cloud (graph $(d)$)
is in agreement with data and identical within one percent to the
ideal Bose gas prediction.
 For the slightly colder cloud of graph $(c)$, the HF
 avoids the divergence in the ideal gas model,
 although it underestimates the peak density by approximately 20\%.
 For the even colder cloud of graph $(b)$, the
discrepancy between the HF profile and the experimental
data is even larger (35\% at the center).

\begin{figure}
\includegraphics{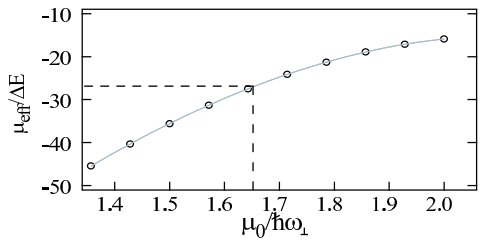}
\caption{ Ratio between the effective chemical potential and
$\Delta E$, the energy splitting between the ground state and the
first excited state as a function of the chemical potential for
the temperature of profile (b). The dashed lines correspond to the
measured chemical potential. 
} \label{figure_ratio}
\end{figure}

 To validate our Hartree Fock calculations we check {\it  a
posteriori} the local density approximation (LDA). The LDA is valid
if the population difference between adjacent energy states is
negligible.
 This criterion is met if
the absolute value of
$\mu_{\rm eff}=
\mu_0-\varepsilon_0(\mu)$ is much larger than $\Delta E$, where
$\varepsilon_0(\mu)$ is the ground state energy and $\Delta E$ the
energy gap between the ground state and the first excited
longitudinal state.
For the temperature of
the graph (b), as long as $\mu_0<2.0 \hbar \omega_{\perp}$,
$|\mu_{\rm eff}/\Delta E|$ is larger than $15$ (see
Fig.~\ref{figure_ratio}),
and for the chemical
potential $\mu_0^b=1.65\hbar\omega_\perp$ deduced from the data
$|\mu_{\rm eff}/\Delta E|\simeq 25$. For such a large value of
$|\mu_{\rm eff}/\Delta E|$, the LDA is expected to be valid.

 For this ratio $\mu_{\rm eff}/\Delta E$,
we can quantify the error made in the density profile due to the
LDA.
 From the HF calculation, we obtain the energies $E_n(\mu,T)$ of
the transverse eigenstates.
 Assuming the transverse motion adiabatically follows
the longitudinal one, we obtain  an effective longitudinal
Hamiltonian with a potential $V_n(z)=E_n(\mu_0-m\omega_z^2z^2/2,T)$
for each transverse mode.
 Diagonalization of each effective longitudinal
Hamiltonian and summation of the resulting thermal profiles gives
the expected longitudinal density profile. We find agreement  with
the profile obtained from the LDA within 5\%. This procedure also
confirms the LDA for the HF calculations corresponding to the graphs
(c) and (d).

 In the case of profile $(b)$, where the gas is in
the quasi-condensate regime at the center,
 the Hartree-Fock
calculation predicts a population of the ground state $N_0\approx
k_B T/(\varepsilon_0-\mu_0)=0.0035 N_{tot}$.
 Therefore, the Hartree-Fock
approach does not predict a saturation of the excited states and
fails to explain the presence of the quasi-condensate at the center
of the cloud. The local density approximation criterion $|\mu_{\rm
eff}|=|\mu_0-\varepsilon_0(\mu)|>>\Delta E$ also implies a small
relative ground state population.

 The failure of the Hartree-Fock approach for our experimental
parameters is due to the large density fluctuations this theory
predicts in a dense, nearly 1D gas.
 When density fluctuations become too large, pair interactions
induce correlations in position between particles which are not
taken into account in the Hartree-Fock theory. These correlations
reduce the interacting energy by decreasing density fluctuations:
the gas enters the quasi-condensate regime.

We now estimate the cross-over density $n_{\rm c.o.}$ at which the
gas enters the quasi-condensate regime. For this purpose, we
assume the gas is in the quasi-condensate regime and use the
Bogoliubov theory to compute density fluctuations. We find {\it a
posteriori} the validity domain of the quasi-condensate regime,
which requires that density fluctuations $\delta \rho$ be small
compared to the mean density $\rho$. More precisely, we define
$n_{\rm c.o.}$ as the density for which the Bogoliubov calculation
yields $\int\!\!\int (\delta\rho(r))^2\!/\!(\rho(r) n_{\rm c.o.})\
d^2r = 1$.
 We indicate this cross-over density in Fig.\ref{figure_resultats}. We
find that $n_{\rm c.o.}$ is close to the density above which the
experimental profiles agree with the quasi-condensate profile.

In conclusion, we have been able to reach a situation where a
quasi-condensate is experimentally observed but a HF approach fails
to explain its presence. As for purely one dimensional systems, the
passage towards quasi-condensate in our experiment is a smooth
cross-over driven by interactions. The profiles that we observe
require a more involved theory able to interpolate between the
classical and the quasi-condensate regime.
 A Quantum Monte Carlo calculation that gives the exact solution
of the many body problem \cite{shiftTcSvistunov,Holzmann1999} should
reproduce the experimental data. In fact, since temperatures are
larger than interaction energy, quantum fluctuations of long
wavelength excitations should be negligible and a simpler classical
field calculation should be sufficient \cite{Goral2002,Davis2006}.
 Finally, the refined mean field theory proposed in \cite{StoofHFmod},
where the coupling constant $g$ is
modified to take into account correlations between atoms,
may explain our profiles.

We thank A. Aspect and L. Sanchez-Palencia for careful reading of
the manuscript, and D.~Mailly from the LPN (Marcoussis, France) for
help in micro-fabrication. The atom optics group is member of
l'Institut Francilien de la Recherche sur les Atomes Froids. This
work has been supported by the EU under grants MRTN-CT-2003-505032,
IP-CT-015714.

\end{document}